\documentclass[preprint,showpacs,amsmath,amssymb,floatfix,pre]{revtex4}

\usepackage{graphicx,color}

\usepackage{graphicx}
\newcommand{\be}{\begin{equation}}
\newcommand{\ee}{\end{equation}}

\begin{document}
\title{Magnetization plateaus in the antiferromagnetic Ising chain with single-ion anisotropy and quenched 
disorder} 

\author{Minos A. Neto}
\email{minosneto@hotmail.com}
\affiliation{Departamento de F\'{\i}sica, Universidade Federal do Amazonas, 3000, Japiim,
69077-000, Manaus-AM, Brazil}

\author{J. Ricardo de Sousa}
\email{jsousa@edu.ufam.br}
\affiliation{Departamento de F\'{\i}sica, Universidade Federal do Amazonas, 3000, Japiim,
69077-000, Manaus-AM, Brazil}
\affiliation{National Institute of Science and Technology for Complex Systems, 3000, Japiim,
69077-000, Manaus-AM, Brazil}
\date{\today}

\author{N. S. Branco}
\email{nsbranco@fisica.ufsc.br}
\affiliation{Departamento de F\'{\i}sica, Universidade Federal de Santa Catarina, 
CEP: 88040-900 Florianópolis-SC}

\begin{abstract}

We have studied the presence of plateaus on the low-temperature magnetization of an antiferromagnetic spin-1 chain, as an external uniform magnetic field is varied. 
A crystal-field interaction is present in the model and the exchange constants follow a random quenched (Bernoulli or Gaussian) distribution. Using a transfer-matrix 
technique we calculate the largest Lyapunov exponent and, from it, the magnetization at low temperatures as a function of the magnetic field, for different values of the 
crystal-field and of the width of the distributions. For the Bernoulli distribution, the number of plateaus increases, with respect to the uniform case (F. Litaiff, J. R. de Sousa, 
and N. S. Branco, Sol. St. Comm. {\bf 147}, 494 (2008)) and their presence can be linked to different ground states, when the magnetic field is varied. For the Gaussian 
distributions, the uniform scenario is maintained, for small widths, but the plateaus structure disappears, as the width increases. 
\end{abstract}

\pacs{75.10.Hk, 75.40.Cx, 75.50.Ee, 75.50.Lk}

\maketitle

\section{Introduction\protect\nolinebreak}

  The physics of one-dimensional or quasi one-dimensional, spin-$S$ ($S\geqslant1$),
chains have attracted a considerable amount of attention since Haldane's prediction \cite{haldane1983} that the one-dimensional 
Heisenberg antiferromagnetic model should have an energy gap between the singlet ground state and the first excited triplet 
states in the case of an integer spin quantum number, while the energy levels are gapless in the case of a half-integer spin quantum number. 

     One fascinating characteristic of some low-dimensional quantum systems is that they 
show magnetic plateaus, i.e., quantization of the magnetization at low temperatures near the ground state. A general 
condition for the quantization of the magnetization in low-dimensional magnetic systems is derived from the 
Lieb-Schultz-Mattis \cite{lieb1961} theorem. One important issue on this matter is the effect of a single-ion anisotropy term on the critical and multicritical 
phenomena, which has been investigated by various theoretical methods, such as, mean-field approximation \cite{liwang1976}, 
transfer-matrix, finite-size-scaling and Monte Carlo simulation \cite{kimel1987,wang1991,plascak2003,lara1998}, Bethe lattice treatment 
\cite{ekiz2004a,ekiz2004b},
and effective field theory \cite{hai2004,yan2002}. The mechanism for the appearance of these magnetization plateaus in quasi 
one-dimensional spin chains are expected to be, among others, dimerization, frustration, single-ion anisotropy, quantum effects, and periodic field. 
However, plateau structures may also be present in classical models at zero temperature. These structures, as we will see, are connected
to different ground-state configurations.

   Experimentally, the magnetization plateaus were observed in high-fields measurements of several magnetic materials 
such as the quasi one-dimensional compounds $SrCu_{2}O_{3}$ \cite{azuma1994}, $Y_{2}BaNiO_{5}$ \cite{darriet1993}, 
$Ni(C_{2}H_{8}N_{2})_{2}NO_{2}ClO_{4}$ (abreviated NENP) \cite{hagiwara1998,narumi1998}, and $Cu(NO_{3})_{2}2.5H_{2}O$ \cite{bonner1983}; 
triangular antiferromagnets $C_{6}Eu$ \cite{sakakibara1983}, $CsCuCl_{3}$ \cite{nojiri1988}, and $RbFe(MoO_{4})_{2}$ 
\cite{inami1996}; and in a quasi two-dimensional compound, with a Shastry-Sutherland lattice structure, $SrCu_{2}(BO_{3})_{2}$ 
\cite{kageyama2000}. The magnetic susceptibility of this compound shows a couple of characteristics of
a spin gap system: a Curie-Weiss behavior at high temperatures, with a round maximum at $20K$, followed by a sudden 
decrease and zero spin susceptibility at $T=0$ \cite{kageyama1999}. Another piece of evidence for the extremely localized nature of triplet excitations is obtained 
when one measures the magnetization curve at low temperatures. Onizuka et al. \cite{onizuka2000} observed a predicted 
plateau at $1/3$ of the total magnetization around $50$ T, in which the magnetic superstructure is characterized 
by a novel stripe order of triplets. He observed that, irrespective of the direction of magnetic field, the
magnetization of $SrCu_{2}(BO_{3})_{2}$ is quantized at $1/8$, $1/4$, and $1/3$ of the fully saturated $Cu$ moment.  

   Our goal in this work is to model the presence and configuration of plateaus in a one-dimensional system with quenched disorder. The presence
of impurities is ubiquitous in nature and the understanding of their effect is, therefore, of great theoretical and experimental relevance. Here we focus
on random and frozen impurities, as opposed to deterministic inhomogeneities (like aperiodic modulation, for example); the latter will be
studied in a forthcoming work.

    This work is organized as follows.  In Section \ref{sec:model} we present the model and formalism. 
In Section \ref{sec:results} the numerical results and discussions are given.  Finally, the last section is devoted to conclusions.

\section{Model and Formalism}  \label{sec:model}

In general, most of the theoretical studies on low-dimensional spin-$S$ systems have been based on Heisenberg 
Hamiltonians, which presents non-trivial quantum effects. However, several studies have shown that it is 
possible to use classical spin systems to obtain magnetic plateaus and study the dependence of these on physical 
characteristics \cite{chen2003,ohanyan2003,aydiner2004,litaiff2008,eloy2011magnetic,qi2012magnetization}, 
as discussed in the Introduction. 
Consider, for example, the Hamiltonian of a one-dimensional spin-$1$ classical system, which 
uses the Ising type variable instead of quantum spin operators:

\begin{eqnarray}
\mathcal{H}=J\sum\limits_{i}S_{i}^{z}S_{i+1}^{z} - H\sum\limits_{i}S_{i}^{z}+ \Delta \sum\limits_{i}(S_{i}^{z})^{2},
\label{eq:hamiltoniana}
\end{eqnarray}%
where $J$ stands for the antiferromagnetic ($J>0$) exchange coupling, $\Delta$ is the single-ion anisotropy
(or the crystal-field term), $H$ is an external uniform magnetic field, and $S_{i}^{z}=\pm 1,0$. 

      In a previous work \cite{litaiff2008}, it has been shown that this model allows for
magnetization plateaus at zero temperature.
The method used to calculate thermodynamic quantities for this Hamiltonian  \cite{litaiff2008}
 was the transfer-matrix technique \cite{kramers1941a,kramers1941b}, which, besides being very simple, 
allows for exact results when applied to one-dimensional classical systems. 
Within this formalism, it can be shown \cite{yeomans1992} that the partition function $Z$ is given, as
a function of the transfer matrix $\mathcal{M}$, by:

\begin{equation}
Z=\textbf{Tr} \; \mathcal{M}^{N} = \lambda_{1}^{N}+\lambda_{2}^{N}+\lambda_{3}^{N}+... 
   \simeq \lambda_1^N,
\label{3}
\end{equation}%
for $N \rightarrow \infty$, where $N$ is the number of spins in a one-dimensional lattice or the number of rows in a two-dimensional one
and $\lambda_1 > \lambda_2 \ge \lambda_3 \ge ...$ are the eigenvalues of  ${\cal M}$.

  Therefore, the free energy per spin reads
\begin{equation}
f(H,T)=-k_{B}T\lim_{N\rightarrow\infty}\frac{1}{N}\ln Z \simeq -k_{B}T\ln \lambda_{1},
\label{eq:energialivre}
\end{equation}%
where $k_B$ is the Boltzmann constant and $T$ is the temperature.

  For the model defined by the Hamiltonian in Eq. (\ref{eq:hamiltoniana}), the transfer matrix reads

\begin{equation}
\mathcal{M}=\left( 
\begin{array}{ccc}
x y z & y z & y z/x \\ 
1 & 1 & 1               \\
y/(x z) & y/z & x y/z 
\end{array}\right) ,
\label{eq:matriz}
\end{equation}%
where $x\equiv\exp(-K)$, $y\equiv\exp(-d)$ and $z\equiv\exp(L)$, with $K=\beta J$, $d=\beta \Delta$, $L=\beta H$, and
$\beta=1/k_B T$.

    Our goal in this work is to study the effects of disorder in the plateaus' structure. We introduce the disorder in the
exchange constant $J$, which will have its value given by a probability distribution (see below).
Therefore, the Hamiltonian now reads
\begin{eqnarray}
\mathcal{H}=\sum\limits_{i} J_i S_{i}^{z}S_{i+1}^{z} - H\sum\limits_{i}S_{i}^{z}+ \Delta \sum\limits_{i}(S_{i}^{z})^{2}, 
\label{eq:hamiltonianadesordenada}
\end{eqnarray}%
such that the transfer matrix is given by
\begin{equation}
\mathcal{M}_i=\left( 
\begin{array}{ccc}
x_i y z & y z & y z/x_i \\ 
1 & 1 & 1               \\
y/(x_i z) & y/z & x_i y/z 
\end{array}\right)
\label{eq:matriz1}
\end{equation}%
with $x_i \equiv\exp(-K_i)$ and $K_i=\beta J_i$ and all other definitions as in Eq. (\ref{eq:matriz}).

   We have used two different probability distributions, namely:
\be
     {\cal P}(J_i) = p \; \delta \left(J_i - J \left(1+\sigma\right)\right) + (1-p) \; \delta \left(J_i - J \left(1-\sigma\right)\right),   \label{eq:distbinomial}
\ee
where $p=1/2$ and $\delta$ is the Dirac delta function, and
\be
    {\cal P}(J_{i})=\frac{1}{\sqrt{2\pi\sigma^{2}}}\exp\left[-\frac{(J_{i}/J-1)^{2}}{2\sigma^{2}}\right].   \label{eq:distgaussiana}
\ee
In both cases, the width of the distribution, in unities of $J$, is given by $\sigma$.

    The expression shown in Eq. (3) is in fact a simplification suitable for uniform systems.
On more general grounds, the partition function is obtained from applying the transfer
matrix to a suitable initial vector a great number of times, such that the dominant behavior is
obtained (this behavior is connected to the greatest eigenvalue of the matrix, for uniform
models). Therefore, the correct expression for the calculation of the free energy per site in
the thermodynamic limit and for a disordered model, within the transfer-matrix framework,
reads:
\be
  \frac{f(H,T)}{k_B T} = -  \Lambda_1 \equiv - \lim_{N \rightarrow \infty} \ln  \left\{ \frac{ \| \prod^N_ {i=1} \mathcal{M}_i \vec{v}_0 \| }{ \| \vec{v}_0 \|} \right\}, \label{eq:lyapunov}
\ee
where now the transfer matrix ${\cal M}_i$ is a stochastic quantity and its elements are taken from a given probability distribution
(see Eq. \ref{eq:matriz1}) \cite{cheung1983equilibrium,glaus1986correlations,glaus1987finite,crisanti1994},
$N$ is the size of the strip, $\vec{v}_0$ is a suitable vector to start the iteration with, and the disordered transfer matrix is applied $N$ times to $\vec{v}_0$. 
This starting vector was varied and the results do not depend on it, for $N \gg 1$. In the previous expression,
$\Lambda_1$ is the greatest Lyapunov exponent \cite{crisanti1994}. It is readily seen that $\Lambda_1$
is the greatest eigenvalue of the transfer matrix for uniform models.    

   Eq. (\ref{eq:lyapunov}) holds for a given realization of the disorder; since we are dealing with a quenched model, the free energy has to be averaged for different realizations of the disorder. This quenched average is represented by $[...]$ and, then, we have
\be
     \frac{f^{ave}}{k_B T} = - [\Lambda_1],     \label{eq:energialivredesordenada1}
\ee
such that the magnetization is given by
\be
     m^{ave} = - \frac{\partial f^{ave}}{\partial H},  \label{eq:magnetizacaodesordenada}
\ee
where the derivative is numerically obtained.

    Some technical points are worth stressing here.  The first relevant point is the size of the chain, $N$, which is supposed
to be big enough such that, after successive applications of the transfer matrix, a given accuracy is obtained for the free energy. For practical purposes, we have iterated Eq. (\ref{eq:lyapunov}) until the
values for the free energy do not differ by more than $0.1 \%$; roughly, this precision is achieved  for $N \gtrsim 10^5$. For this value of $N$, the number of disorder
realizations, $N_d$, which lead to an error of $1 \%$ or less in the magnetization is $N_d \sim 10$. The error in $m^{ave}$ is obtained as one standard deviation 
among the $N_d$ values. For the numerical derivatives involved in 
Eq. (\ref{eq:magnetizacaodesordenada}), a value of $\delta H$ (in units of $J$) of order $10^{-4}$ was used. Another 
important point is that our calculation is made at low temperatures.
Therefore, the interaction constant $K$ is big and the iterated vectors may have big components. To avoid overflow,
we normalize the vectors after each iteration of the transfer matrix and use their norms to calculate the free energy, using Eq. (\ref{eq:lyapunov}).
Note also that, for finite temperatures, the free energy is an analytic function of the field. This analyticity precludes mathematically flat
regions, with sharp transitions. However, on the scale of measurements or on numerical calculations, such plateau
regions can and do occur. Moreover, an extrapolation procedure to the zero-temperature
limit clearly shows that the usual plateau structure, with sharp transitions between steps, is obtained.

\section{Results and Discussion}  \label{sec:results}

\begin{figure}[htbp]
\centering
\includegraphics[width=7.5cm,height=7.5cm]{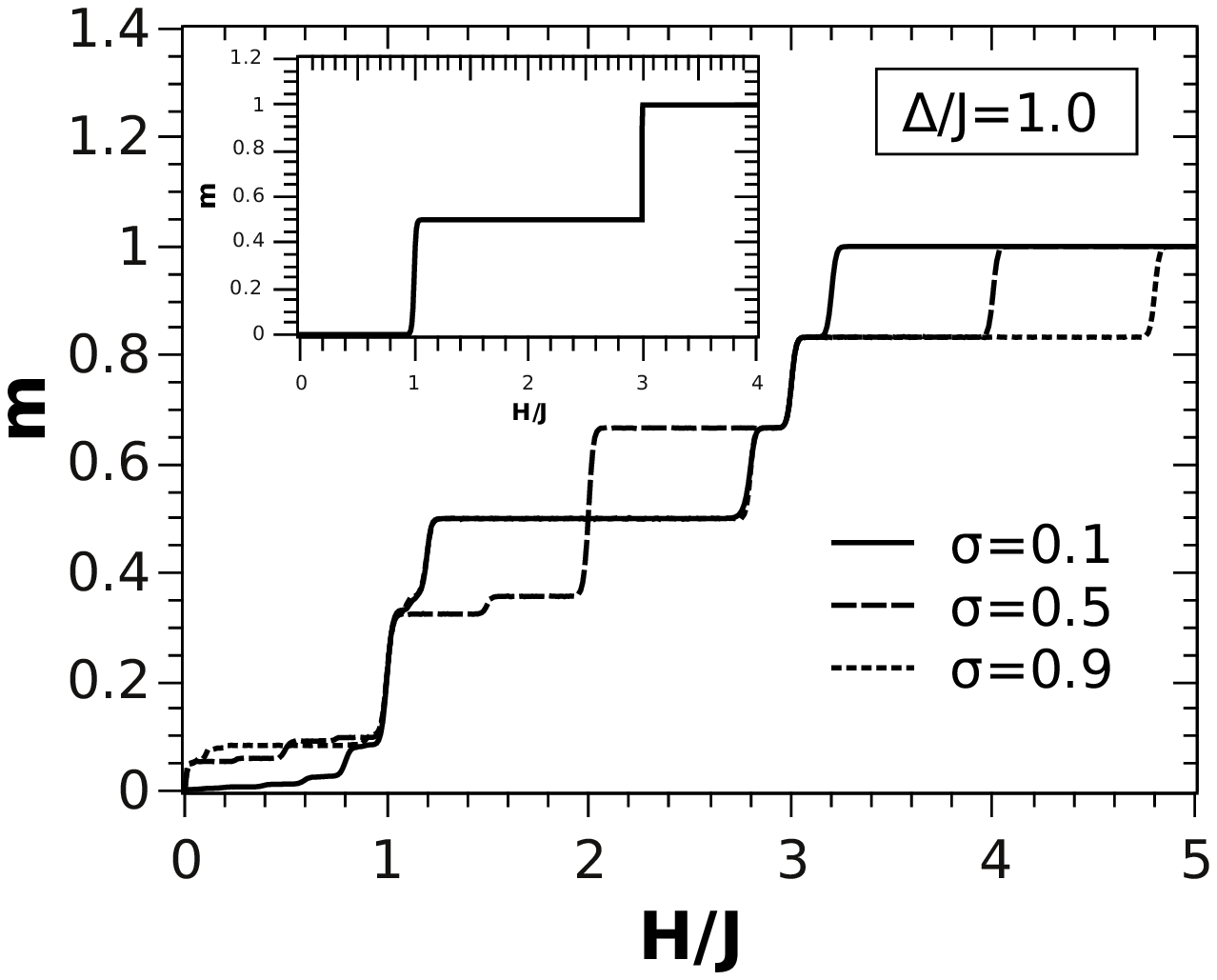}
\caption{Low-temperature magnetization plateaus for the disordered model defined by the Hamiltonian in Eq. (\ref{eq:hamiltonianadesordenada}), with
$\Delta/J=1$. The exchange constant
is obtained from a Bernoulli distribution (Eq.  (\ref{eq:distbinomial})) with width $\sigma$. In the inset, we show the plateau structure
for the uniform model, for comparison.} 
\label{fig:distbinomial} 
\end{figure}

     First, we show our results for the uniform case, in order to test our procedure and code. The results are presented in the inset of
Fig. \ref{fig:distbinomial}, where the magnetization $m$ is depicted as a
function of the reduced magnetic field, $H/J$, for $\Delta/J=1$: they agree with those obtained in Refs. 
\onlinecite{chen2003} and \onlinecite{litaiff2008}. In fact, since our procedure is less time- and memory-consuming, we are able to go to 
temperatures closer to zero and there is less rounding in our plateaus than for the ones in Ref. \onlinecite{chen2003}. 

       For $0 \leq \Delta/J \leq 1$, the first plateau (see inset of Fig. \ref{fig:distbinomial}), at $m=0$, is linked to the configuration $(+-)$, 
the second one, at $m=1/2$, reflects the
configuration $(+0)$, and the last one is connected to the configuration $(++)$. Here, $+$, $-$, and $0$ stand for the values $+1$, $-1$, and $0$ for the spins,
respectively.  The actual configuration of the whole chain is obtained by repeating the two-spin structure inside
parentheses. The first step happens at $H/J=2-\Delta/J$ and the second one takes place at $H/J=2+\Delta/J$.
On the other hand, for $\Delta/J \geq 1$, the first plateau, with $m=0$, is connected to the configuration $(00)$, while the other two plateaus correspond
to the same configurations as for $0 \leq \Delta/J \leq 1$. The step at the left-hand side now happens at $H/J=\Delta/J$ and the step at the right-hand side is
placed in the same position as for $0 \leq \Delta/J \leq 1$. We have also calculated the magnetization vs magnetic field behavior for $\Delta/J < 0$: only two
plateaus are present, since negative values of $\Delta$ do not favor the zero state  \cite{litaiff2008}.

  We now present our results for the disordered model given by the
Bernoulli distribution (see Eq. (\ref{eq:distbinomial})). In Fig. \ref{fig:distbinomial}, where the plateaus structure is shown
for three different values of the width $\sigma$ and for $\Delta/J=1$ (the qualitative results do not differ for other values of $\Delta/J \geq 0$; for
$\Delta/J \leq 0$, as stated above, there are only two plateaus for the uniform model but the introduction of disorder brings the same overall structural changes as for the
three-plateau configuration), for a temperature $T=0.01$ (henceforth, we refer to the temperature in unities of $J/k_B$, i.e., $k_B T/J$).
We notice that the number of plateaus increases with the disorder but we still can connect them to different ground-state configurations.

\begin{figure}[htbp]
\centering
\includegraphics[width=11.5cm]{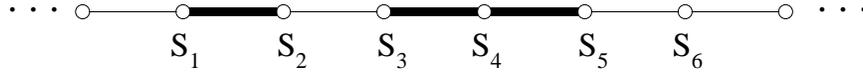}
\caption{In this figure we represent a possible portion of the one-dimensional lattice. Weak interactions 
are represented by thin lines (like the one between spins $S_2$ and
$S_3$, for example) and strong interactions are represented by thick lines (like the one between spins $S_3$ and $S_4$,
for example).} 
\label{fig:rede} 
\end{figure}

   In order to make this connection, we depict, in Fig. \ref{fig:rede}, a portion of the chain, with weak and strong interactions
represented by thin and thick lines, respectively. Due to the random nature of the interactions spins may be 
connected to their two first-neighbors by weak interactions only, by strong ones only or by one of each. In what follows,
we show some examples that link the steps in Fig \ref{fig:distbinomial} with ground-state configurations.

\begin{figure}[htbp]
\centering
\includegraphics[width=11.5cm]{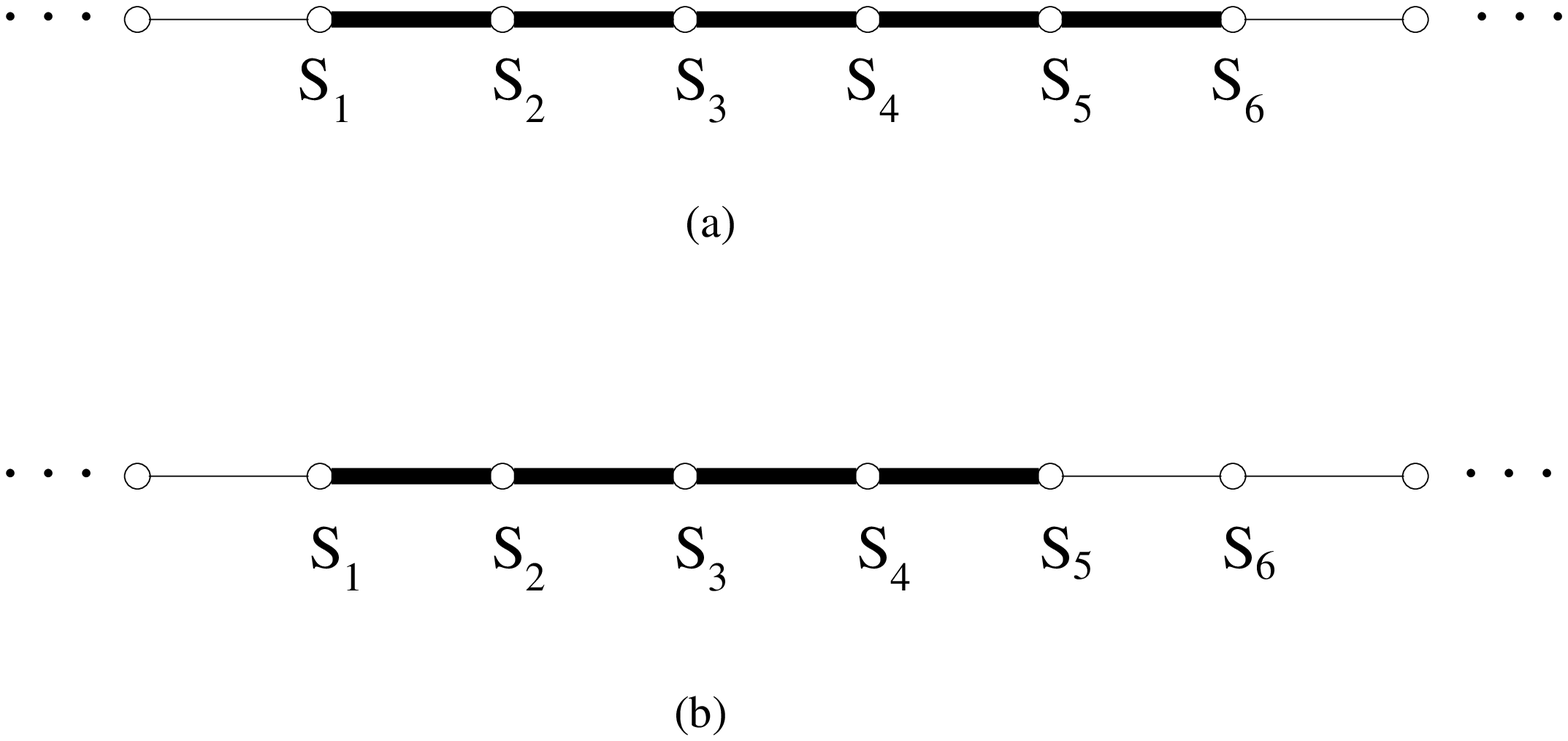}
\caption{Here we depict clusters of spins connected by strong interactions with an even number of spins $(a)$
and an odd number of spins $(b)$. In the former, spins $S_2$ and $S_4$ (or $S_3$ and $S_5$) will 
change first to the $0$ state when $H$ is decreased. In the latter, spins $S_2$ and $S_4$ will turn first under the same
conditions.}
\label{fig:cluster1} 
\end{figure}

   Let us imagine that the magnetic field is strong enough that all spins are in the $+1$ state; this corresponds to $m=1$
(see Fig. \ref{fig:distbinomial}). When the field is decreased,
spins connected to their two first neighbors by strong links will change first, to the $0$ state. But not all spins with this
neighborhood will turn, since these changes increase the contribution of the magnetic-field
term to the energy (although decreasing the corresponding contribution of the exchange and
crystal-field terms to the Hamiltonian). It is
energetically favorable to turn every other spin in a cluster of contiguous strong interactions.
In Fig. \ref{fig:cluster1} $(a)$, spins represented by $S_2$ and $S_4$ (or by $S_3$ and $S_5$) will turn first, when $H$ is decreased, while
in Fig. \ref{fig:cluster1} $(b)$ spins $S_2$ and $S_4$ will go to the $0$ state when the magnetic field is decreased. In any case,
the quoted spins will turn when the magnetic field is decreased and crosses the critical value $H_c$:
\be
 H_c/J \equiv 2 \left( 1 + \sigma \right) + \Delta/J ,  \label{eq:locfirststep}
\ee
which correspond to the three rightmost steps in Fig. \ref{fig:distbinomial} (recall that $\Delta/J=1$).
To calculate the drop in the magnetization on
these steps, one has to work out the statistics of random islands of bonds, which is a one-dimensional percolation problem \cite{aharony1994}. 
The probability of a cluster of $s$ strong bonds to be generated is $p^s (1-p)^2$, while the number of spins that change
state is $n_s=s/2$, for even $s$, and $n_s=(s-1)/2$, for odd $s$. Therefore, the fraction of spins that change state, $\varphi_s$, is:
\be
     \varphi_s = \sum_{s=2, s \; even}^{\infty} \frac{s}{2} p^s (1-p)^2 +  \sum_{s=3, s \; odd}^{\infty}  \frac{(s-1)}{2} p^s (1-p)^2,
\ee
 where the first sum is over even values of $s$ and the second one over odd values of $s$. For $p=1/2$, we obtain
 $\varphi_s=1/6=0.166....$. Therefore, the magnetization drops to $5/6=0.833...$, independent of $\sigma$.

\begin{figure}[htbp]
\centering
\includegraphics[width=11.5cm]{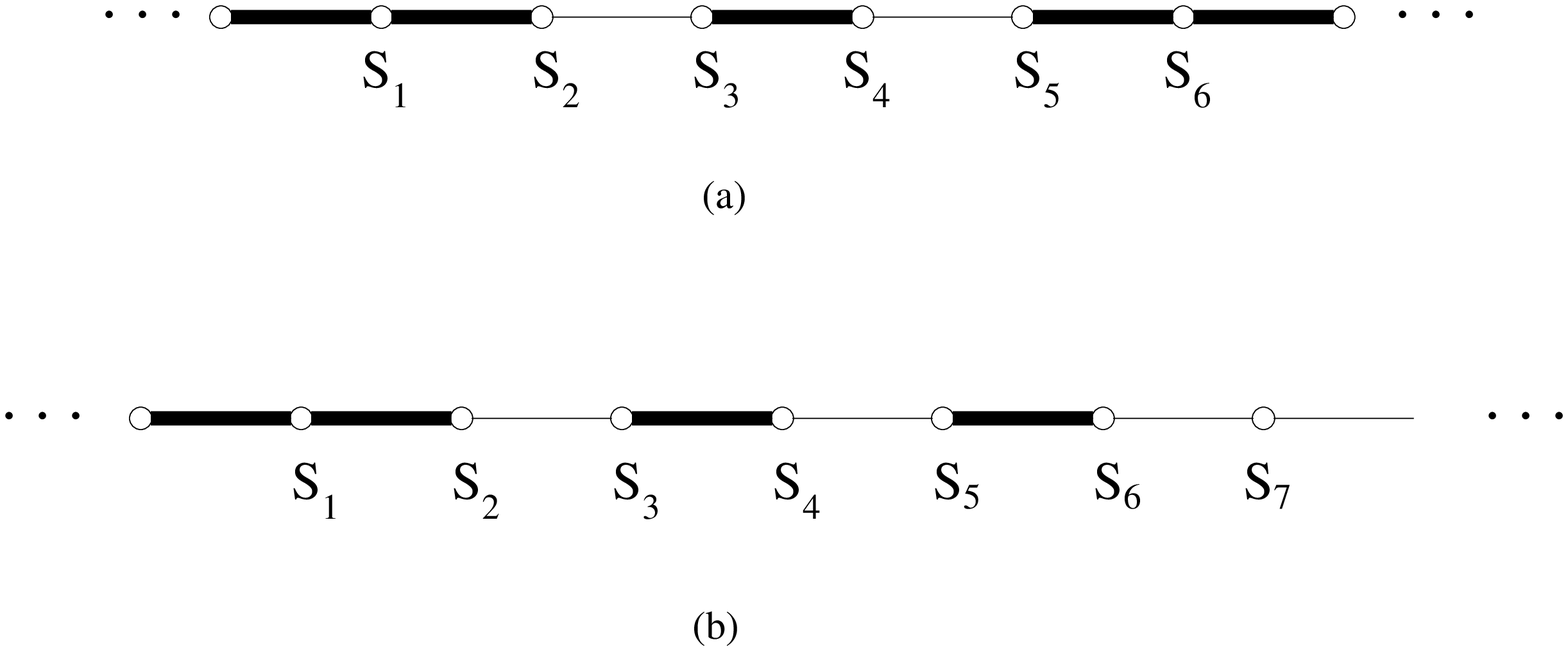}
\caption{Here we depict clusters of spins which interact with
their nearest neighbors via one strong and one weak bond with an even number of spins $(a)$
and with an odd number of spins $(b)$. In the former, spins $S_2$ and $S_4$ (or $S_3$ and $S_5$) will 
change first to the $0$ state when $H$ is decreased. In the latter, spins $S_2$, $S_4$, and $S_6$ will turn first under the same
conditions.}
\label{fig:cluster2} 
\end{figure}

     At $H/J=3$ another step is present, now independent of $\sigma$, and the drop in $m$ is again $1/6$. The spins that change 
state are those in clusters such that the spins interact with
their nearest neighbors via one strong and one weak bond, like the ones depicted in Fig. \ref{fig:cluster2}. 
The spins which change state are, for example, $S_2$ and $S_4$ (or $S_3$ and $S_5$)
in Fig. \ref{fig:cluster2} $(a)$ and $S_2$, $S_4$, and $S_6$ in Fig. \ref{fig:cluster2} $(b)$. The calculation is the same as for the preceding case
and, therefore, the drop on the magnetization is also $1/6$. However, the energy balance is a little bit different and the step
takes place at
\be
   H_c/J = 2 +  \Delta/J.
\ee

   The next step is connected to spins on islands of contiguous weak bonds and the process is the same as the one described 
above for islands of strong contiguous bonds. The magnetization decreases by $1/6$ and the steps are located at:
\be
   H_c/J = 2 ( 1 - \sigma ) + \Delta/J
\ee

   Equivalent steps are obtained when spins on the $0$ state turn to the $-1$ state. These steps are placed on smaller values of $H/J$.

   Therefore, the presence of random disorder described by a Bernoulli distribution increases the total number of steps,
when compared to the uniform model, for the same valeu of $\Delta/J$. However, the steps
are still connected to the energy balance between different configurations of ground states (see Fig. \ref{fig:distbinomial}).
   
\begin{figure}[htbp]
\centering
\includegraphics[width=7.5cm,height=7.5cm]{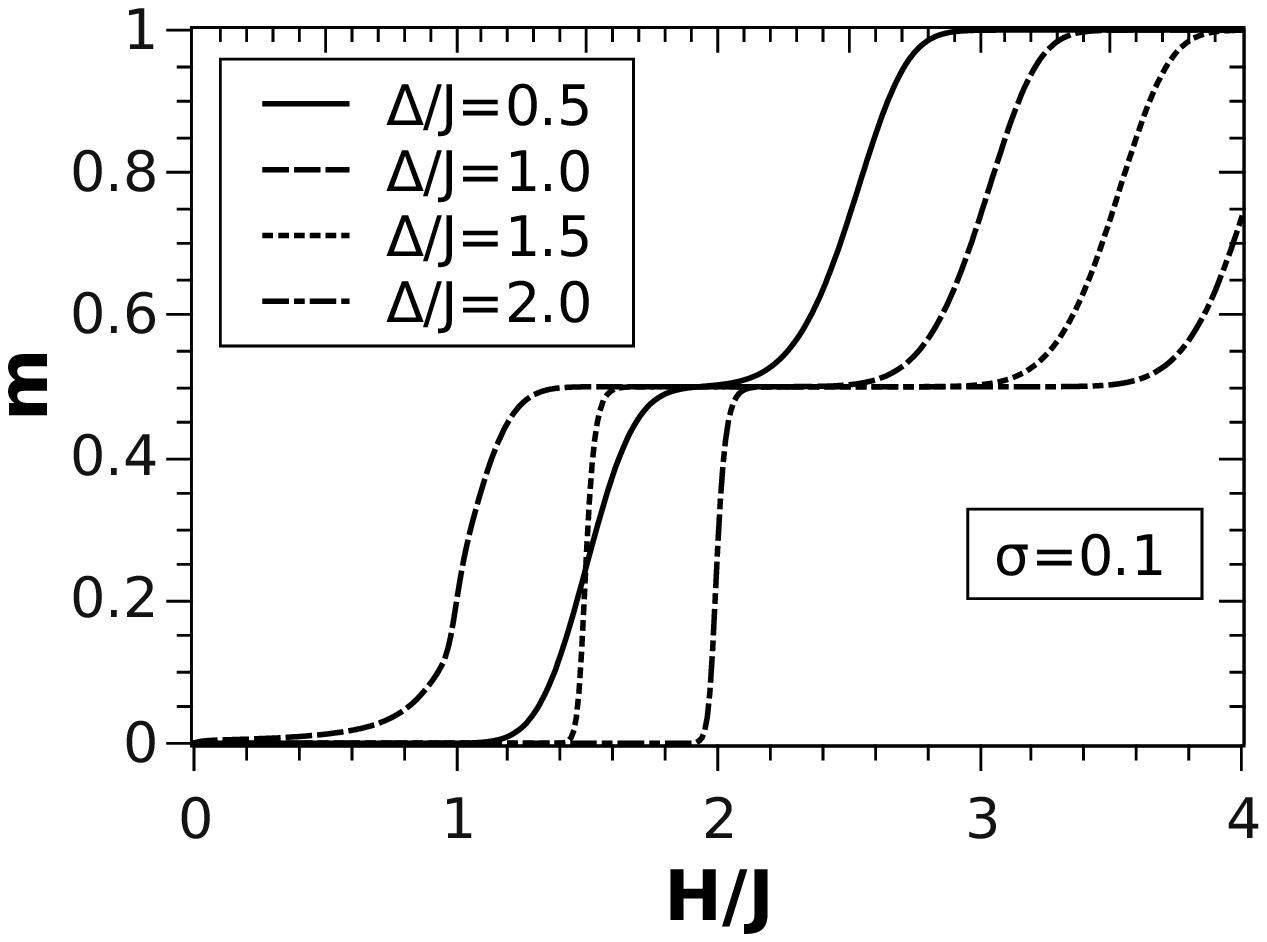}
\caption{Magnetization versus magnetic field for different anisotropy parameters $\Delta/J$, with variance $\sigma=0.1$, and
at $T=0.01$, for the Gaussian distribution, Eq. (\ref{eq:distgaussiana}).}
\label{sigma1} 
\end{figure}

\begin{figure}[htbp]
\centering
\includegraphics[width=7.5cm,height=7.5cm]{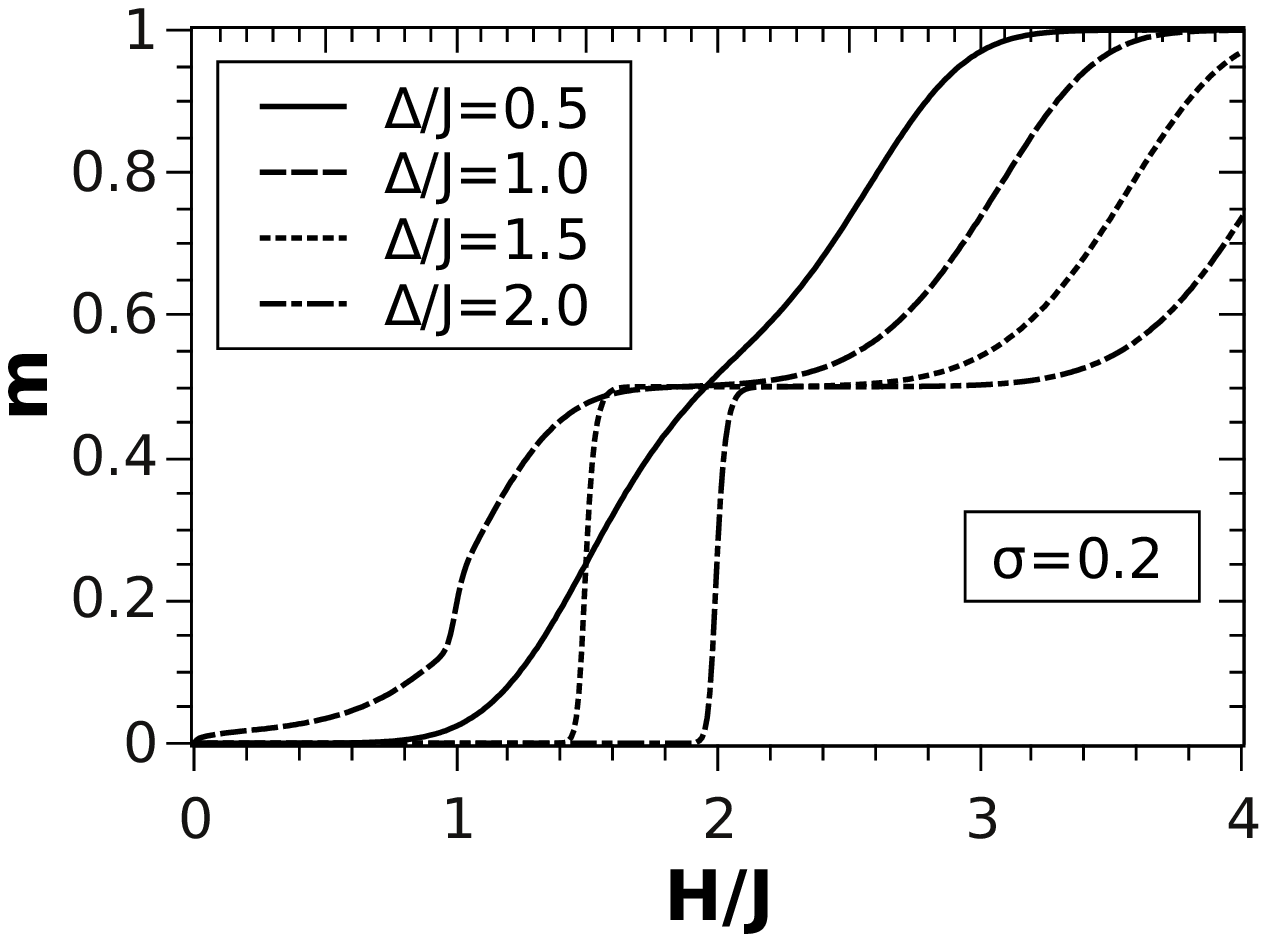}
\caption{Magnetization versus magnetic field for different anisotropy parameters $\Delta/J$, with variance $\sigma=0.2$, and
at $T=0.01$, for the Gaussian distribution, Eq. (\ref{eq:distgaussiana}).} 
\label{sigma2}
\end{figure}

   We now turn to a Gaussian distribution. We calculate the magnetization as a function of 
 $\Delta /J$, the magnetic field $H$ and the variance 
$\sigma$ of the Gaussian probability distribution given by Eq. (\ref{eq:distgaussiana}).
The numerical results are obtained as follows: in Figs. \ref{sigma1} and \ref{sigma2}, the 
magnetization $m$ is plotted at $T=0.01$ 
as a function of $H/J$ and for four values of the anisotropy parameter 
(namely, $\Delta/J=$ $0.5$, $1.0$, $1.5$, and $2.0$), with $\sigma=0.1$ and $\sigma=0.2$ respectively. 

   As expected (see Fig. \ref{sigma1}), the results for low $\sigma$ are qualitatively equivalent to those for the uniform model and
three plateaus are present for $\Delta/J > 0$. This equivalence comes from the fact that, for small values of $\sigma$, the Gaussian distribution 
is concentrated around the value $J_i=J$ (see Eq. (\ref{eq:distgaussiana})). This mimics the uniform model, where all exchange
interactions are the same. For the Bernoulli distribution, on the other hand, no matter 
the (finite) width of the distribution, the two peaks have the same height. Therefore,
two different exchange parameters will be involved in this model. This behavior is depicted
in Fig. \ref{fig:distbinomial}: as long as $\sigma \neq 0$, there will be more steps than for the uniform model. But
we can see that some steps will disappear when $\sigma$ goes to 0, regaining the uniform model
behavior. Note, however, that the width of the plateau diminishes when disorder is introduced, with the
corresponding rounding of the steps. This is consistent with the interpretation that the steps are connected to transitions between different ground states.
The continuous probability distribution leads to a variety of strengths for the exchange interaction, such that different spins change state at different
values of $H/J$. Note that the steps are still centerd at the values for the corresponding steps for the uniform model (as discussed in the first paragraph
of Sec. \ref{sec:results}); the effect of the distribution is only to smooth some of these steps. Also, the left-hand side steps for $\Delta/J = 1.5$ and $2.0$ are not rounded;
the reason is that the energy balance for the transition between states $(00)$ and $(+0)$ (see again first paragraph of Sec. \ref{sec:results}) does not involve the 
exchange constant.

      The picture discussed in the last paragraph is supported by the results depicted in Fig. \ref{sigma2}, for $\sigma = 0.2$. A  wider distribution for $J$ allows for a
 more pronounced rounding of some steps (the left-hand steps for $\Delta/J=1.5$ and $2.0$ remain as for the uniform model, as discussed). 
 Another consequence of this picture is that, for $\Delta/J=0.5$, the intermediate plateau is not present anymore (compare with Fig. \ref{sigma1}, where this 
 plateau, for $\Delta/J=0.5$, still exists, although it is narrower than for $\Delta/J \geq 1$).
 
\begin{figure}[htbp]
\centering
\includegraphics[width=7.5cm,height=7.5cm]{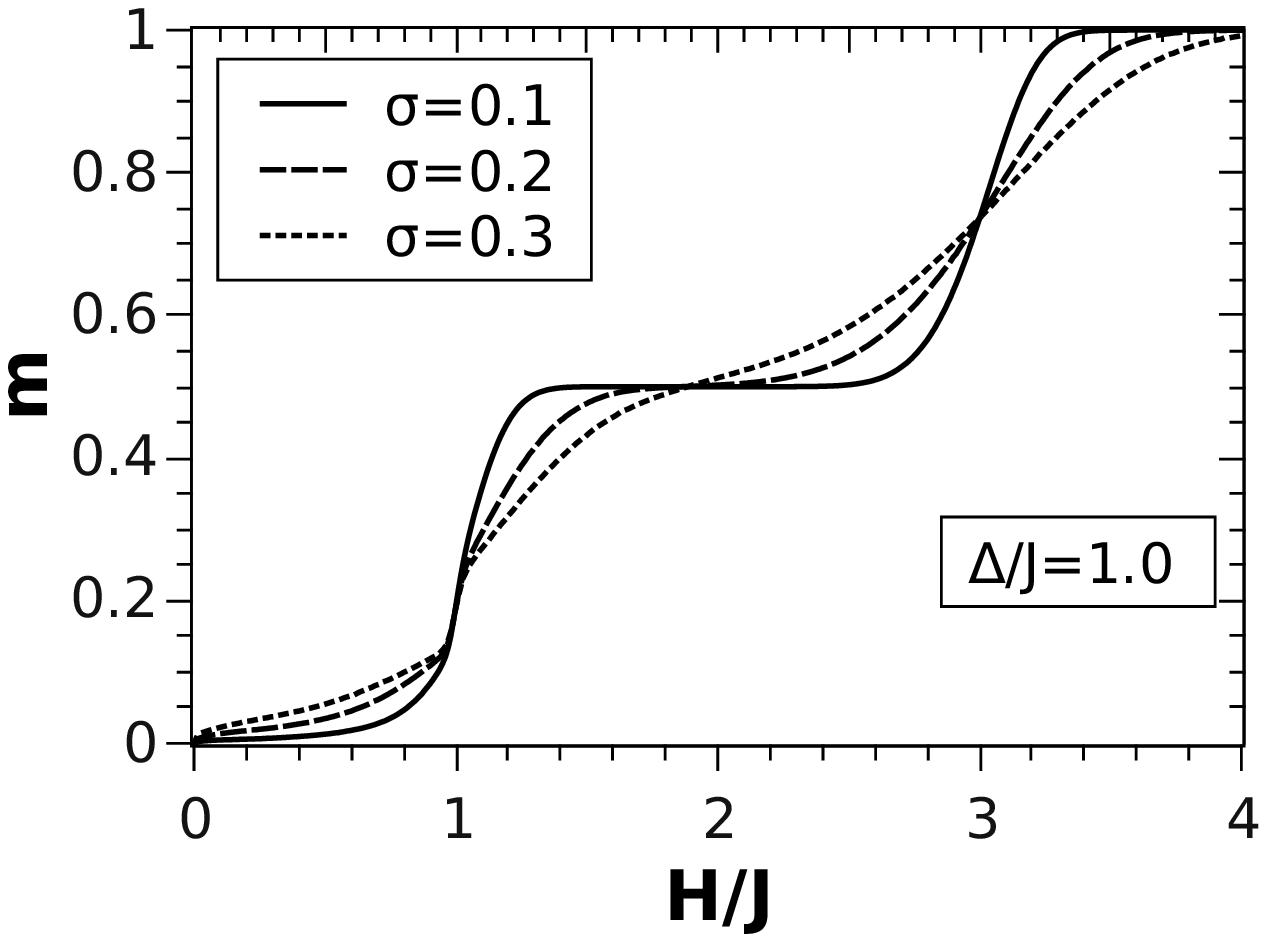}
\caption{Magnetization versus magnetic field for three different values of $\sigma$, with anisotropy parameter $\Delta/J=1.0$, and 
at $T=0.01$, for the Gaussian distribution, Eq. (\ref{eq:distgaussiana}).} 
\label{delta3}
\end{figure}

\begin{figure}[htbp]
\centering
\includegraphics[width=7.5cm,height=7.5cm]{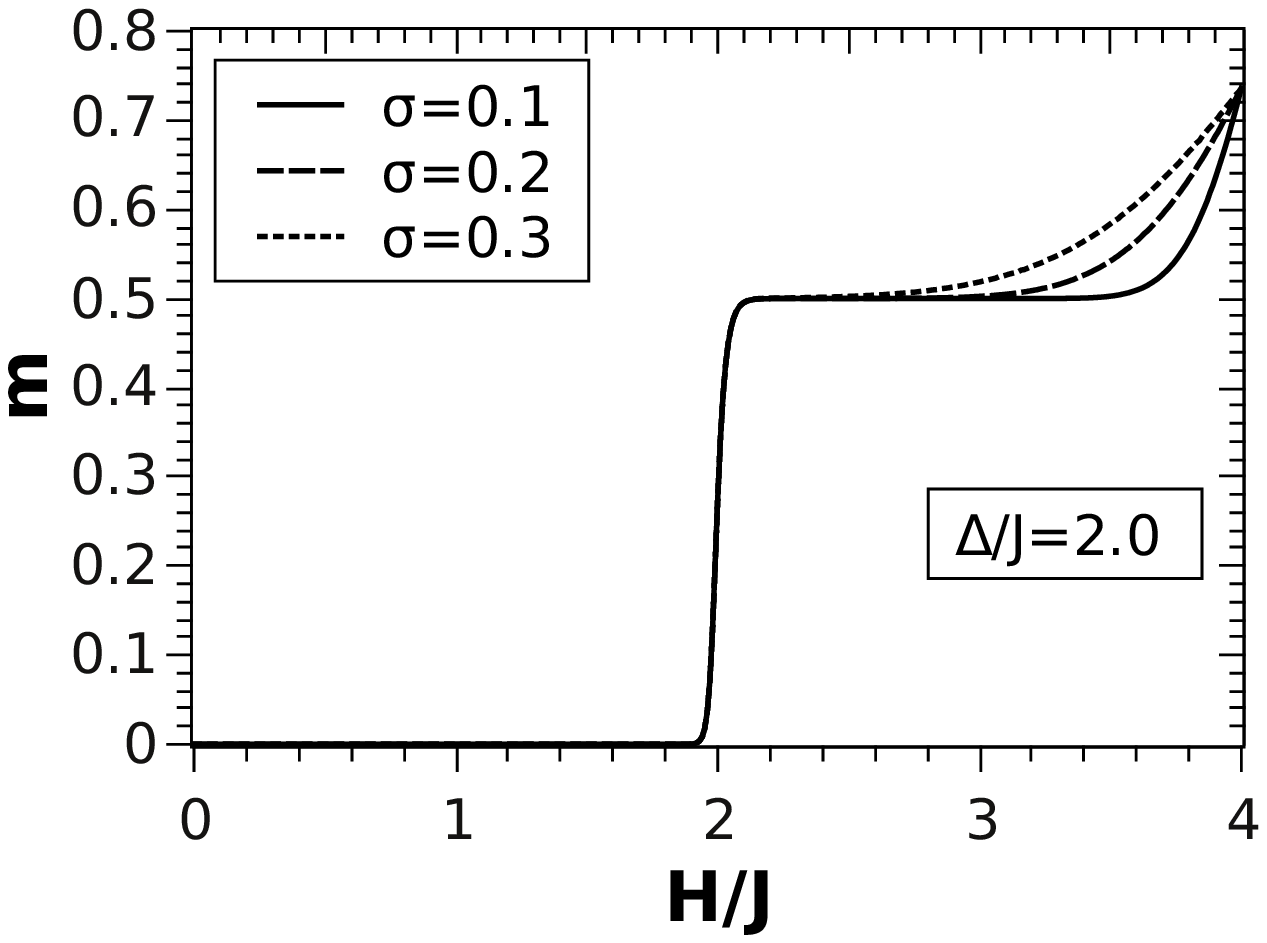}
\caption{Magnetization versus magnetic field for three different values of $\sigma$, with anisotropy parameter $\Delta/J=2.0$, and 
at $T=0.01$, for the Gaussian distribution, Eq. (\ref{eq:distgaussiana}).}
\label{delta5}
\end{figure}

    In Figs. \ref{delta3}  and \ref{delta5}, the magnetization  $m$ at $T=0.01$  is plotted as a function of  $H/J$ for several values of the variance $\sigma$ ($0.1$, $0.2$,
and $0.3$) with  $\Delta/J=1.0$ and $2.0$, respectively. These figures show that the steps are rounded when the width of the Gaussian distribution is increased.

  Our results indicate that the Gaussian distribution only rounds the steps for small values of its width. For large ones, the
continuous distribution tends to eliminate the steps and plateaus. This is
consistent with the observation that these plateaus are consequence of transitions between different spin configurations at the ground state.

   The presence of magnetic plateaus in the present system confirms that this structure may be obtained in classical models. 
Similar behavior has been observed in theoretical studies \cite{chen2003,aydiner2005} and in experimental results \cite{narumi1998}.
Note that the ground 
state phase diagrams are symmetric with respect to $H/J=0$ line, i.e. invariant under the transformation $H\rightarrow-H$ 
and $S\rightarrow-S$. As a final note, we would like to mention that the effects of disorder depend on the model and may be completely different from the ones observed
in this work. When applied to a quantum model, for example, disorder following a Gaussian distribution completely eliminates the presence of plateaus
\cite{cabra2000random}.

\section{Conclusions}

In this work, we studied the low temperature magnetization plateaus of a classical spin system with discrete symmetry. 
By using a transfer-matrix technique, we calculate the magnetization behavior at low temperatures 
as a function of the anisotropy parameter $\Delta/J$, magnetic field $H$, and variance $\sigma$ of the
probability distributions. It is found that positive single-ion anisotropies have a significant effect on the low-temperature 
magnetic properties in the system, as the external field is varied. The number of plateaus for the Bernoulli distribution increases, when
compared to the uniform model, and their presence is linked to transitions between different ground states.
For the Gaussian distribution and small width $\sigma$, only one intermediate plateau (at $m=1/2$)
is found. The steps in the magnetization-versus-magnetic field graphs are rounded as the width of the Gaussian 
distribution is increased and
the overall picture is consistent with the presence of steps due to transitions between different
ground states. Note, however, that the effects of the introduction of a Gaussian distribution depends strongly on the underlying model
(see, for example, Ref. \onlinecite{cabra2000random}).
Our results are compatible with previous experimental and theoretical studies. Work is under way on
spin-alternating chain models, which are expected to show many interesting phenomena \cite{tonegawa2010}.

\begin{acknowledgments}
We thank Octavio Daniel Salmon and I. J. L. Diaz
for useful discussions. This work was partially supported by PPP/FAPEAM-012/2009, FAPESC,
CAPES, and CNPq (Brazilian Research Agencies).
\end{acknowledgments}

\bibliography{references}

\newpage


\vspace{1.0cm}

\vspace{1.0cm}

\vspace{1.0cm}

\vspace{1.0cm}

\vspace{1.0cm}

\vspace{1.0cm}



\vspace{1.0cm}


\vspace{1.0cm}

\end{document}